\documentclass{PoS}
\usepackage{graphicx}
\usepackage{graphics}
\usepackage{epsfig}
\usepackage{lineno}
\newcommand{\be}{\begin{eqnarray}}
\newcommand{\ee}{\end{eqnarray}}
\newcommand{\ave}[1]{\left\langle #1 \right\rangle}

\newcommand{\SPD}{\rm{SPD}}

\newcommand{\VZERO}{\rm{VZERO}}

\newcommand{\nudyn}{$\nu_{(+-,dyn)}$}

\newcommand{\deltaeta}{$\Delta \eta$~}
\newcommand{\nchnudyn}{$N_{ch}\times\nu_{(+-,dyn)}$}
\title{Charge fluctuations in Pb--Pb collisions at $\sqrt{s_{NN}} = $2.76 TeV measured by ALICE experiment}

\ShortTitle{Charge fluctuations in Pb--Pb collisions at $\sqrt(s_{NN}) = $ 
  2.76 TeV measured by ALICE experiment}

\author{\speaker{Satyajit Jena}%
  ~for the ALICE\thanks{A Large Ion Collider Experiment, LHC-CERN}~~Collaboration \\
  Indian Institute of Technology Bombay, 
  Powai-400 076, Mumbai, India\\
  E-mail: \email{sjena@cern.ch}}

\abstract{Charge fluctuations provide a possible signature for the 
  existence of the de-confined Quark Gluon Plasma phase (QGP). 
  Being sensitive to the square of the charges, fluctuations
  in QGP, with fractionally charged partons, are significantly 
  different from those of a hadron gas with unit charged particles. 
  Studies of charge fluctuations have been carried out  using 
  the variable, $\nu_{(+-,dyn)}$ which, by its construction, is 
  free from collisional bias (impact parameter fluctuations 
  and fluctuations from the finite number of charged particles within 
  the detector acceptance). The dependence of charge fluctuations 
  on the pseudo--rapidity windows for various centrality bins is analyzed 
  for Pb--Pb collisions at $\sqrt{s_{\rm NN}}$~=~2.76~TeV in the ALICE 
  experiment at CERN-LHC. A scaling behavior is observed as a function 
  of increasing pseudo-rapidity window for the charge fluctuations, 
  expressed in terms of \nchnudyn, where $N_{ch}$ is the number of 
  charged particles. The results are compared to experimental 
  measurements at lower energies and to model predictions.}

\FullConference{The Seventh Workshop on Particle Correlations and 
  Femtoscopy\\
  September 20 - 24 2011\\
  University of Tokyo, Japan}

\begin{document}

\section{Introduction}
Heavy-ion collisions at ultra-relativistic energies can produce 
a new state of matter characterized by high temperature 
and energy density, where the degrees of freedom are given not
by hadrons but by their constituents, the quarks and gluons~\cite{satz}. 
The ALICE experiment~\cite{Aamodt08}, located at the CERN LHC, is a 
multi-purpose experiment with highly sensitive detectors around the 
interaction point. The central detectors cover the pseudo-rapidity  
region $|\eta| < 0.9$, with good momentum measurement as well as good 
impact parameter resolution. This gives us an excellent 
opportunity to study the fluctuations and correlations of physical 
observable on an event--by--event basis. Details of the ALICE experiment 
and its detectors may be found in \cite{Aamodt08}. 

The fluctuations of net--charge depend on the squares of the charge 
states present in the system. The QGP phase, having  quarks as the 
charge carriers, should result in a significantly different magnitude 
of fluctuation compared to a hadron gas (HG). As discussed in~\cite{JeonKoch99, JeonKoch00}, 
The net--charge fluctuation is measured in terms of D defined as
\be 
D = 4 {\ave{\delta Q^2} \over \ave{N_{\rm ch}}}
 \label{eq:Dpm}
 \ee
where $Q = N_+ - N_-$ is the net--charge and $N_{\rm ch} = N_+ + N_-$, here $N_+$ and $N_-$ 
are the numbers of positive and negative particles.
The net--charge fluctuation expressed in term of $D$ is predicted to be 4 
for non--interacting pion-gas, $\simeq$3 for hadron resonance gas and 
$\simeq$1-1.5 for QGP~\cite{JeonKoch1}. 

However, on 
an event--by--event basis the fluctuations are best studied experimentally
through ``non-statistical'' or ``dynamical'' fluctuations.
The dynamic charge observable, \nudyn is defined as 
\begin{eqnarray}
\nu_{(+-,dyn)} = \nu _{ +  - } - \nu _{ stat} = 
 \frac{\langle N_+(N_+-1) \rangle}{\langle N_+ \rangle ^2} +
 \frac{\langle N_-(N_--1) \rangle}{\langle N_- \rangle ^2} 
- 2\frac{\langle N_-N_+ \rangle}{\langle N_- \rangle \langle N_+ \rangle},
\end{eqnarray} 
where
\begin{equation}
\nu _{ +  - }  = \left\langle {\left( {\frac{{N_ +  }}{{\left\langle {N_ +  } 
\right\rangle }} - \frac{{N_ -  }}{{\left\langle {N_ -  } \right\rangle }}} 
\right)^2 } \right\rangle 
\end{equation}
\noindent and 
\begin{equation}
\nu _{ stat}= \frac{1}{{\left\langle {N_ +  } \right\rangle }} + 
\frac{1}{{\left\langle {N_ -  } \right\rangle }}
\end{equation}
\noindent and $\langle .... \rangle$ denotes the average over all events. 
And the \nudyn~is a measure of the relative correlation~\cite{css} strength 
of $++$, $--$, 
and $+-$ particles pairs. Note that by construction, these correlations 
are identically zero for Poissonian, or independent particle production. 
Furthermore  $D$ can be expressed in terms of \nudyn~as
\begin{eqnarray}
  D \approx \nu_{(+-,dyn)}  \times \langle N_{\rm ch} \rangle + 4
\end{eqnarray}

In this article we report the first measurement of the net--charge 
fluctuations in Pb--Pb collisions at  $\sqrt{s_{NN}}=2.76$~TeV measured 
with the ALICE detector.The data were recorded in November 2010 during 
the first run with heavy ions at the LHC. In this analysis, the Time 
Projection Chamber (TPC)~\cite{TPC} is used 
for selecting tracks, the Inner Tracking System (ITS) is used for vertexing 
and triggering and the VZERO scintillator detector is used for estimating 
centrality~\cite{toia} as well as triggering. The collision centrality 
is determined by cuts on the \VZERO~multiplicity as described in~\cite{ALICEcharged}.  
A study based on Glauber model fits \cite{Glauber} to the multiplicity 
distribution in the region corresponding to thw 90\% of most central 
collisions, where the vertex reconstruction is fully efficient, 
facilitates the determination of the cross section percentile and the 
number of participants. The resolution in centrality is found to 
be $< 0.5$\% RMS for the most central collisions (0-5\%), increasing 
towards 2\% RMS for peripheral collisions (70-80\%).
The present analysis is performed by taking vertexes within $\pm10$ cm from the 
nominal interaction point along the beam axis (z) to ensure a uniform 
acceptance in the central pseudo-rapidity $|\eta| < 0.8$ and the charged 
particle transverse momentum, 
$p_{\rm T}$, from 0.15 GeV/$c$ to 5 GeV/$c$. The trigger consisted of a hit on  
the two \VZERO~scintillator detectors, positioned on both sides of the 
interaction point, in coincidence with a signal from the Silicon Pixel 
Detector (\SPD). We have removed background events using the \VZERO~timing 
information and the requirement of at least two tracks in the central detectors.

\begin{figure}[htb!]
\begin{center}
\includegraphics[scale=0.6]{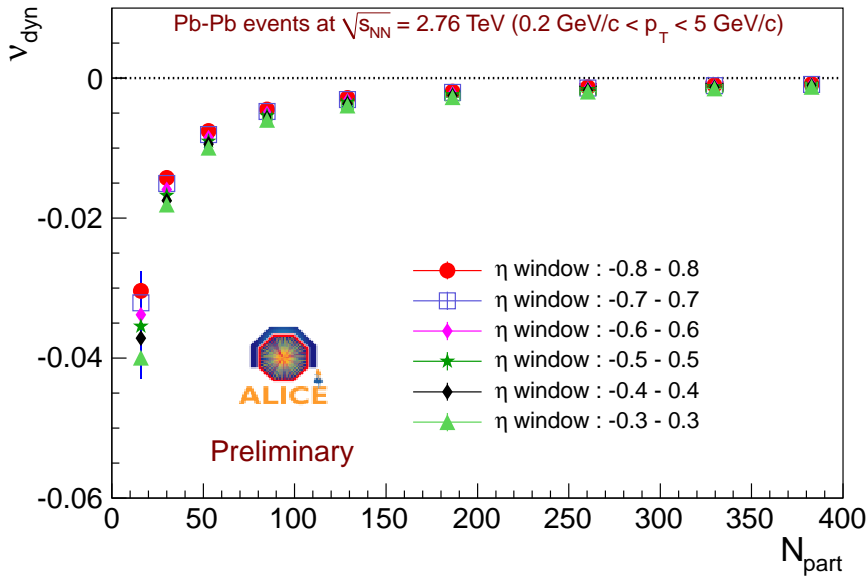}
\caption{Dynamical net--charge fluctuations, \nudyn, of charge 
particles within different pseudo-rapidity  windows, as a function of number 
of participants.} 
\label{fig1}
\vskip -8mm
\end{center}
\end{figure}

The contribution to the systematic uncertainty originating from the following 
were considered: (a) uncertainty in the determination of interaction 
vertex, (b) the effect of magnetic field, (c) contamination from secondary 	
tracks (DCA cuts), (d) centrality definition using different detectors, 
and (e) quality cuts of the tracks. The systematic and statistical 
uncertainties in the plots are represented by the shaded areas and the error 
bars, respectively.

The dynamic fluctuations, \nudyn, were calculated on an event--by--event basis 
from the measurements of positive and negatively charged particles 
produced within \deltaeta windows defined around mid-rapidity. Fig.~\ref{fig1} 
shows, the \nudyn~as a
function of N$_{part}$, where 
moving from left to right along the x-axis implies moving from the most central 
to the most peripheral collisions. The value of \nudyn~decreases monotonically, 
going from central to peripheral collisions for various \deltaeta windows.

\begin{figure}[htb!]
\begin{center}
\includegraphics[scale=0.6]{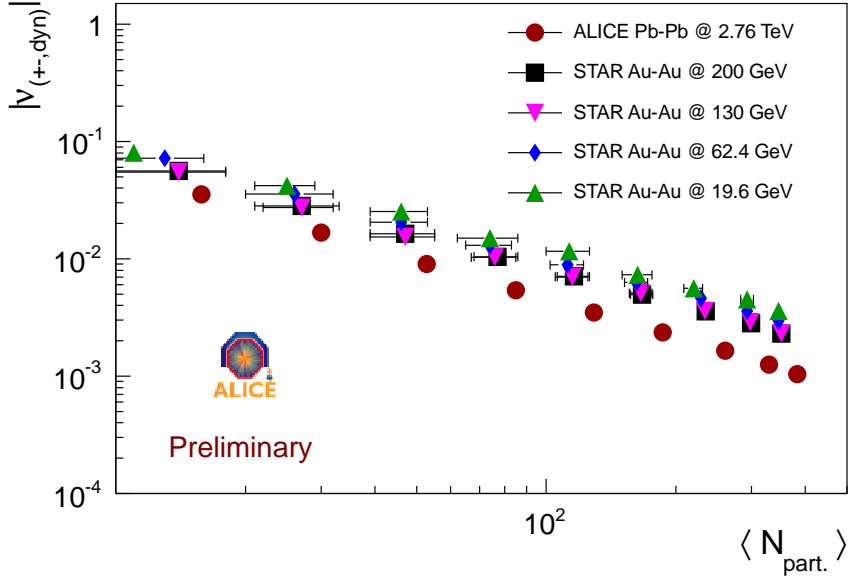}
\caption{The absolute value of \nudyn,  as a function of 
 the collision centrality compared with measurements for lower energies.}
\label{fig2}
\vskip -8mm
\end{center}
\end{figure}

We have studied the beam energy dependence of the net--charge fluctuations 
by combining the ALICE points with those of RHIC data~\cite{star}. In Fig.~\ref{fig2}, 
we present the absolute value \nudyn~as a function of number of
participants for \deltaeta$=1 $, in  \mbox{Pb--Pb} collisions at $\sqrt{s_{NN}}=2.76$~TeV 
at LHC and Au--Au collisions at STAR. In all cases  dynamical net--charge fluctuations 
exhibit a monotonic dependence on the number of participating nucleons.
The ALICE data are below the STAR points for Au--Au collisions 
at all centralities, indicating an additional reduction of the magnitude 
of fluctuations at LHC energies.

\begin{figure}[htb!]
  \centering
\includegraphics[scale=0.6]{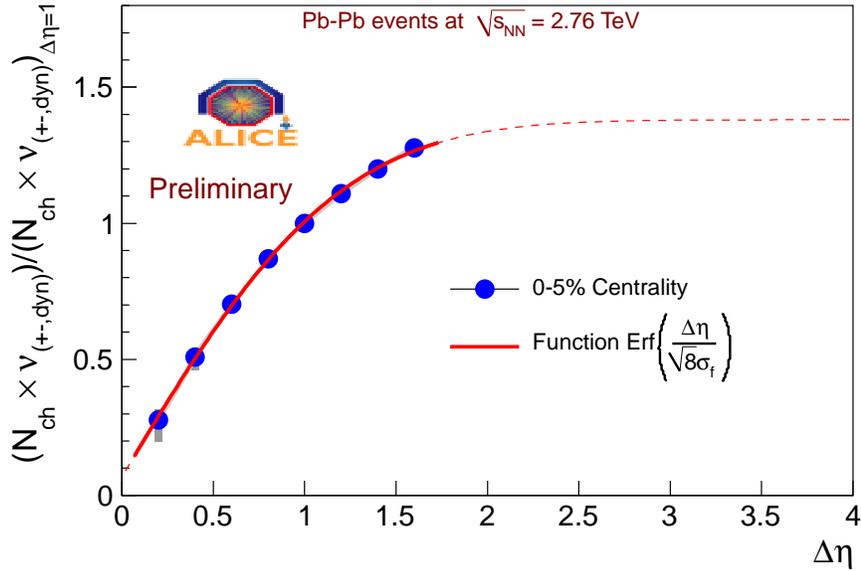}
\vskip -0.2cm
\caption{${\rm N}_{ch}\times\nu_{(+-,dyn)}$, normalized to the values 
for \deltaeta$=1$, are plotted as a function of $\Delta \eta$. The data points 
are fitted with the functional form erf(\deltaeta$/\sqrt{8}\sigma_f$) 
normally used for diffusion equations. The dashed line is an 
extrapolation of the fitted line.\label{fig3} } 
\end{figure}

\begin{figure}[htb!]
\centering
\includegraphics[scale=0.6]{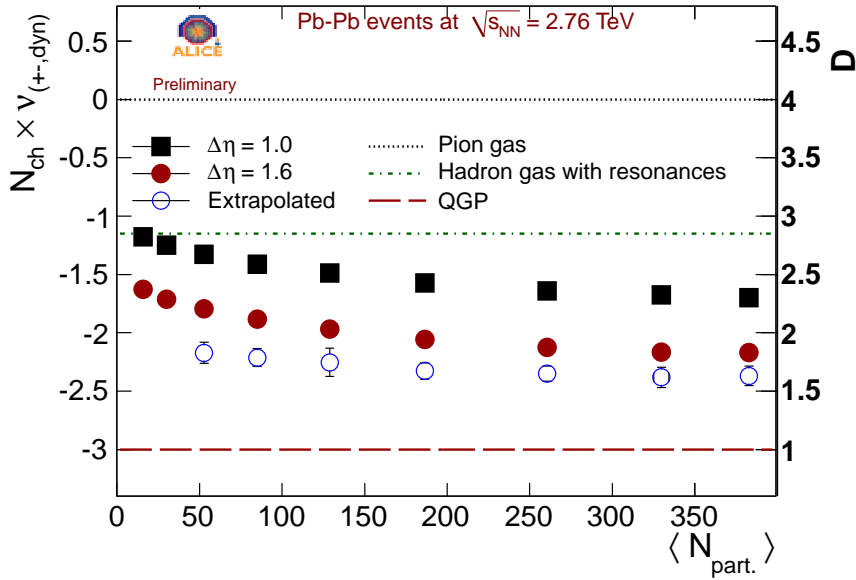}
\vskip -0.2cm
\caption{\nchnudyn~(left-axis) and D (right-axis) are plotted for \deltaeta$=1$ and 1.6 as 
  a function of the number of participants. The values after extrapolating to higher 
  \deltaeta are shown by open circles. \label{fig4} } 
\end{figure}

We examine the nature of the variation of \nchnudyn~with \deltaeta by plotting its 
ratio with respect to the value at \deltaeta$=1$, as shown in Fig. 3. 
We observe that the relative value of \nchnudyn 
~ grows smoothly with
increasing \deltaeta window. This behavior has been predicted in 
\cite{Shuryak01,AbdelAziz05} and was attributed to the spread
of the signal arising from the diffusion during the evolution 
from the early QGP stage to the hadron resonance gas (HG). Charge conservation and longitudinal 
expansion affect the growth, which may limit the increase to an asymptotic 
value. We fit the data points of Fig.~\ref{fig3} with a function of the form 
erf(\deltaeta$/\sqrt{8}\sigma_f$)~\cite{sg,sg1}, 
representing the diffusion in rapidity space, where $\sigma_f$ is the diffusion 
parameter. The diffusion coefficient, $\sigma_f$, obtained from the fitting is equal 
to $ 0.467 \pm 0.02$ at 0-5\% centrality. An extrapolation of the fitted 
value indicates the onset of saturation at \deltaeta$=3$. 
It has been conjectured that, taking only dissipation into account, the asymptotic 
value of fluctuations may give back the original value of fluctuations at the 
early QGP stage.

In Fig.~\ref{fig4}, the net--charge 
fluctuations, expressed in terms of \nchnudyn~and $D$ (left-- and 
right--axis, respectively) as a 
function of the $N_{part}$ are shown for three different 
\deltaeta windows, i.e. $\Delta \eta = 1$, $\Delta \eta = 1.6$ and the 
extrapolated asymptotic values at $\Delta \eta = 3$, along with the lines 
indicating the predicted values of fluctuations for three cases: pion gas, 
HG and QGP. The values at 
asymptotic limits are obtained for each centrality bin, separately. 
A decreasing trend of fluctuations is observed, measured in terms 
of D, when going from peripheral to central collisions. By confronting 
the measured value with the theoretically predicted fluctuations 
\cite{JeonKoch00, Shuryak01}, it is observed that the results are 
within the limits of the QGP and the HG scenarios. 
In reality the fluctuation might have been less than 
the observed value, because of further damping due to the final state 
interactions, expansion, collision dynamics, string fusion, or other effects 
discussed in ~\cite{sg,sg1,str1,str2,str3,str4}.

In summary, we have presented the first measurements of dynamic net--charge 
fluctuations at the LHC in \mbox{Pb--Pb} collisions at $\sqrt{s_{NN}}=2.76$~TeV 
using the observable $\nu_{(+-,dyn)}$. The net--charge fluctuations are observed 
to be dominated by the correlations of oppositely charged particles. 
The energy dependence of the dynamical fluctuations shows a decrease 
in fluctuation going from RHIC to LHC energies. A fit to the fluctuation in rapidity 
space is using the diffusion equation, which yields the asymptotic 
value of fluctuation, which is closer to the theoretically predicted 
value of Quark Gluon Plasma.

\end{document}